\documentstyle[psfig,rotate]{mn}
\input{psfig}
\newif\ifAMStwofonts
\ifoldfss

  \ifCUPmtlplainloaded \else
    \NewTextAlphabet{textbfit} {cmbxti10} {}
    \NewTextAlphabet{textbfss} {cmssbx10} {}
    \NewMathAlphabet{mathbfit} {cmbxti10} {}
    \NewMathAlphabet{mathbfss} {cmssbx10} {}
  \fi
  \ifAMStwofonts
    \ifCUPmtlplainloaded \else
      \NewSymbolFont{upmath} {eurm10}
      \NewSymbolFont{AMSa} {msam10}
      \NewMathSymbol{\upi}     {0}{upmath}{19}
      \NewMathSymbol{\umu}     {0}{upmath}{16}
      \NewMathSymbol{\upartial}{0}{upmath}{40}
      \NewMathSymbol{\leqslant}{3}{AMSa}{36}
      \NewMathSymbol{\geqslant}{3}{AMSa}{3E}

       \let\le=\leqslant
       
    \fi
  \fi
\fi 
\ifnfssone
  \newmathalphabet{\mathit}
  \addtoversion{normal}{\mathit}{cmr}{m}{it}
  \addtoversion{bold}{\mathit}{cmr}{bx}{it}

  \newmathalphabet{\mathbfit} 
  \addtoversion{normal}{\mathbfit}{cmr}{bx}{it}
  \addtoversion{bold}{\mathbfit}{cmr}{bx}{it}
  \newmathalphabet{\mathbfss} 
  \addtoversion{normal}{\mathbfss}{cmss}{bx}{n}
  \addtoversion{bold}{\mathbfss}{cmss}{bx}{n}
  \ifAMStwofonts
    \ifCUPmtlplainloaded \else
      \UseAMStwoboldmath
      \makeatletter
      \new@mathgroup\upmath@group
      \define@mathgroup\mv@normal\upmath@group{eur}{m}{n}
      \define@mathgroup\mv@bold\upmath@group{eur}{b}{n}
      \edef\UPM{\hexnumber\upmath@group}
      \new@mathgroup\amsa@group
      \define@mathgroup\mv@normal\amsa@group{msa}{m}{n}
      \define@mathgroup\mv@bold\amsa@group{msa}{m}{n}
      \edef\AMSa{\hexnumber\amsa@group}
      \makeatother
      \mathchardef\upi="0\UPM19
      \mathchardef\umu="0\UPM16
      \mathchardef\upartial="0\UPM40
      \mathchardef\leqslant="3\AMSa36
      \mathchardef\geqslant="3\AMSa3E

       \let\le=\leqslant
       
    \fi
  \fi
\fi 
\ifnfsstwo

  \DeclareMathAlphabet{\mathbfit}{OT1}{cmr}{bx}{it}
  \SetMathAlphabet\mathbfit{bold}{OT1}{cmr}{bx}{it}
  \DeclareMathAlphabet{\mathbfss}{OT1}{cmss}{bx}{n}
  \SetMathAlphabet\mathbfss{bold}{OT1}{cmss}{bx}{n}
  \ifAMStwofonts
    \ifCUPmtlplainloaded \else
      \DeclareSymbolFont{UPM}{U}{eur}{m}{n}
      \SetSymbolFont{UPM}{bold}{U}{eur}{b}{n}
      \DeclareSymbolFont{AMSa}{U}{msa}{m}{n}
      \DeclareMathSymbol{\upi}{0}{UPM}{"19}
      \DeclareMathSymbol{\umu}{0}{UPM}{"16}
      \DeclareMathSymbol{\upartial}{0}{UPM}{"40}
      \DeclareMathSymbol{\leqslant}{3}{AMSa}{"36}
      \DeclareMathSymbol{\geqslant}{3}{AMSa}{"3E}

       \let\le=\leqslant
       
    \fi
  \fi
\fi 
\ifCUPmtlplainloaded \else
  \ifAMStwofonts \else 
    \def\upi{\pi}
    \def\umu{\mu}
    \def\upartial{\partial}
  \fi
\fi
\title{Measuring Cold Dark Matter Power Spectrum
from Variations of Hubble Flows}
\author[Xiangdong Shi]
   {Xiangdong Shi\\
   Department of Physics, University of California, La Jolla, CA 92093, USA}
\date{Accepted 
      Received 
      in original form}
\pubyear{1997}
\begin{document}
\maketitle
\begin{abstract}
When Cold Dark Matter (CDM) power spectrum normalized by COBE results,
its amplitude at smaller scales can be parametrized by $\Gamma\sim\Omega_0h$.
The expected variations of Hubble flows in two samples,
the sample of 36 clusters in the Mark III catalogue, and the sample of 20 Type
Ia supernovae (SNe), are calculated for the power spectrum in
critical-density CDM models (including tilted CDM models and vacuum-dominated
CDM models).  The comparison between the expectations and the
real variations in the data offers a bias-free way to constrain the power
spectrum.  The cluster sample yields $\Gamma\le 0.30-0.88(n_{\rm ps}-1)+1.9
(n_{\rm ps}-1)^2$ at $95\%$ C.L., with best fits $\Gamma =
0.15-0.39(n_{\rm ps}-1)+0.37(n_{\rm ps}-1)^2$, where $n_{\rm ps}$ is
the spectral index of the power spectrum.
The Type Ia SN sample yields $\Gamma\le 0.25-0.80(n_{\rm ps}-1)+1.6
(n_{\rm ps}-1)^2$ at $95\%$ C.L., strongly favoring lower $\Gamma$'s.  The
results are inconsistent with a critical-density matter-dominated universe with
$n_{\rm ps}\ga 0.8$ and $H_0\ga 50$ km/sec/Mpc.
\end{abstract}
\section{Introduction}

Variations in Hubble flows (peculiar Hubble flows,
or monopolar deviations from a global Hubble flow)
are directly tied to the underlying density
fluctuation.  For example, for a sphere with a uniform density embedded in a
homogeneous FRW universe with a critical density,
its peculiar Hubble flow under linear theory is
\begin{equation}
{\delta H\over H_0}=-{1\over 3}{\delta\rho\over \rho},
\end{equation}
where $\delta H/H_0$ is the local deviation from a global Hubble flow
normalized by the  Hubble constant $H_0$, and $\delta\rho/\rho$
is the local density contrast.
For realistic $\Omega_0=1$ models with gaussian power spectra,
the r.m.s. peculiar Hubble flow as measured within a top-hat
sphere is, however (Turner, Cen and Ostriker 1992; Shi, Widrow and Dursi 1995)
\begin{equation}
\Bigl\langle\Bigl({\delta H\over H_0}\Bigr)^2\Bigr\rangle^{1\over 2}
\approx -0.6\Bigl\langle\Bigl({\delta M\over M}\Bigr)^2\Bigr\rangle^{1\over 2}.
\end{equation}
Thus, if we know the variations of Hubble flows as a function of
scales, we have a grasp of the power spectrum that generates the underlying
density fluctuations.  Since only peculiar velocities are involved,
the approach is free of the bias factor $b$.  And ideally, it will not be
influenced by structures beyond the sample volume.

Conventionally, the power spectrum $P(k)$ in critical-density CDM models
(including tilted CDM models and $\Lambda$CDM Models where
$\Lambda$ is the cosmological constant) is parametrized by
\begin{equation}
P(k)=2\pi^2\delta ^2(3000h^{-1}{\rm Mpc})^{3+n_{\rm ps}}k^{n_{\rm ps}}T^2,
\end{equation}
where (Bunn and White 1997)
\begin{equation}
\delta =1.94\times 10^{-5}\Omega_0^{-0.785-0.05\ln\Omega_0}
        e^{-0.95(n_{\rm ps}-1)-0.169(n_{\rm ps}-1)^2},
\label{normalize}
\end{equation}
and (Bardeen et al. 1986; Sugiyama 1995)
\begin{eqnarray}
T&=&{\ln(1+2.34q)\over 2.34q
[1+3.89q+(16.1q)^2+(5.46q)^3+(6.71q)^4]^{1/4}},\nonumber\\
q&=&k/\Gamma h,\quad \Gamma=\Omega_0 he^{-\Omega_b-\sqrt{h/0.5}
\Omega_b/\Omega_0}.
\end{eqnarray}
In the equations, $\Omega_0$ is the matter content of the universe,
$\Omega_b$ (adopted to be $0.024h^{-2}$, Tytler, Fan and Burles 1996)
is the baryonic content of the universe, and $h$
is the Hubble constant in the unit of 100 km/sec/Mpc.  Eq. (\ref{normalize})
is due to normalization by the COBE results.  Spectral index $n_{\rm ps}$ is
constrained to between $\approx 0.8$ and 1.1 (Lineweaver and Barbosa 1997).
Given the COBE normalization, $\Gamma$ and $n_{\rm ps}$ determine the shape
of the power spectrum.

In this article we calculate the expected variations of Hubble flows
for real samples using the above form of the CDM power spectrum.  We then
compare the expectations to real variations and obtain
constraints on $\Gamma$ as a function of $n_{\rm ps}$.

\section{Formalism}
A detailed formalism can be found in Shi (1997).  Here we only present some
essential equations.

The deviation from a global Hubble flow, $\delta H$
($=H_{\rm Local}-H_0$), of a sample and its bulk motion
${\bf U}$ are 
\begin{equation}
\delta H=B^{-1}\sum_q{S_qr_q-U_ir_q\hat r_q^i\over\sigma_q^2},
\label{pechub}
\end{equation}
\begin{equation}
U_i=(A-RB^{-1})^{-1}_{ij}\Bigl(\sum_q{S_q{\hat r}_q^j\over\sigma_q^2}-
    B^{-1}\sum_q\sum_{q^\prime}{S_qr_qr_{q^\prime}{\hat r}_{q^\prime}^j\over
    \sigma_q^2\sigma_{q^\prime}^2}\Bigr),
\label{bulkmotion}
\end{equation}
where
\begin{equation}
A_{ij}=\sum_q{\hat r_q^i\hat r_q^j\over\sigma_q^2},\,\,
R_{ij}=\sum_q\sum_{q^\prime}{r_q\hat r_q^ir_{q^\prime}
\hat r_{q^\prime}^j\over\sigma_q^2\sigma_{q^\prime}^2},\,\,
B=\sum_q{r_q^2\over\sigma_q^2}.
\label{Aij}
\end{equation}
Vector ${\bf r}_q$ is the position of object $q$ in the sample (with
us at the origin), and
$S_q$ is its estimated line-of-sight peculiar velocity with
an uncertainty $\sigma_q$.

The expectation of the deviation $\delta H$ consists of two
parts: the contribution from the density fluctuations $\delta H^{(v)}$,
and the contribution from noises $\delta H^{(\epsilon)}$.  

In linear theory,
\begin{equation}
\Bigl\langle\Bigl({\delta H^{(v)}\over H_0}\Bigr)^2\Bigr\rangle
=\Omega_0^{1.2}\int d^3k
\bigl\vert W^i({\bf k})\hat k_i\bigr\vert ^2{P(k)\over k^2},
\end{equation}
where the window function of the sample
\begin{eqnarray}
W^i({\bf k})&=&{B^{-1}\over (2\pi)^{3/2}}
   \Bigl[\sum_q {r_q^i\over\sigma_q^2}e^{i{\bf k}\cdot{\bf r}_q}\nonumber\\
&-&(A-RB^{-1})^{-1}_{jl}\Bigl(\sum_q{\hat r_q^i\hat r_q^j\over\sigma_q^2}
   e^{i{\bf k}\cdot{\bf r}_q}\nonumber\\
&-&B^{-1}\sum_q\sum_{q^\prime}{r_q^ir_{q^\prime}^j\over
   \sigma_q^2\sigma_{q^\prime}^2}e^{i{\bf k}\cdot{\bf r}_q}\Bigr)
   \sum_{q^{\prime\prime}}{r_{q^{\prime\prime}}^l
   \over\sigma_{q^{\prime\prime}}^2}\Bigr].
\end{eqnarray}
Spatial indices $i,j,l,m$ run from 1 to 3.  Identical indices denote
summation.

The r.m.s. noise contribution is
\begin{equation}
\Bigl\langle\Bigl({\delta H^{(\epsilon)}\over H_0}\Bigr)^2\Bigr\rangle
^{1\over 2}=B^{-1}+B^{-2}(A-RB^{-1})^{-1}_{il}R_{il}.
\label{noiseself}
\end{equation}

Since we do not know the precise value of $H_0$, we can only deal with
relative variations of Hubble flows within a sample.
If we denote the expansion rate defined by a subsample with
$n$ nearest objects (relative to us) as $H_n$,
that by another subsample with $m(<n)$
nearest objects as $H_m$, and that by the entire sample with $N$ objects as
$H_N$, then the expected
\begin{eqnarray}
&&\Bigl\langle\Bigl({H_n-H_m\over H_N}\Bigr)^2\Bigr\rangle\nonumber\\
&\approx&\Bigl\langle\Bigl({\delta H_n^{(v)}\over H_0}\Bigr)^2
+\Bigl({\delta H_m^{(v)}\over H_0}\Bigr)^2-2\Bigl({\delta H_n^{(v)}\over H_0}
 \Bigr)\Bigl({\delta H_m^{(v)}\over H_0}\Bigr)\Bigr\rangle\nonumber\\
&+&\Bigl\langle\Bigl({\delta H_n^{(\epsilon)}\over H_0}\Bigr)^2
+\Bigl({\delta H_m^{(\epsilon)}\over H_0}\Bigr)^2
-2\Bigl({\delta H_n^{(\epsilon)}\over H_0}\Bigr)
\Bigl({\delta H_m^{(\epsilon)}\over H_0}\Bigr)\Bigr\rangle,
\label{variation}
\end{eqnarray}
in which
\begin{eqnarray}
&&\Bigl\langle\Bigl({\delta H_n^{(v)}\over H_0}
\Bigr)\Bigl({\delta H_m^{(v)}\over H_0}\Bigr)\Bigr\rangle\nonumber\\
&=&\Omega_0^{1.2}\int d^3k\,
W^i_n({\bf k})\hat k_i{W^j_m}^*({\bf k})\hat k_j\, {P(k)\over k^2}
\end{eqnarray}
and
\begin{equation}
\Bigl\langle\Bigl({\delta H_n^{(\epsilon)}\over H_0}\Bigr)
\Bigl({\delta H_m^{(\epsilon)}\over H_0}\Bigr)\Bigr\rangle=
\Bigl\langle\Bigl({\delta H_n^{(\epsilon)}\over H_0}\Bigr)^2\Bigr\rangle.
\label{noisecross}
\end{equation}

To better account for the contribution from fluctuations on very small
scales at which the window functions are choppy and random (see figure 1)
and density fluctuations become non-linear, we use
the variation of $\delta H$ as a measure of the amplitude of
density fluctuations, and represent the small-scale contribution
by a one-dimension gaussian random motion with a r.m.s. velocity
$\sigma_*$.  The expectation is then
\begin{eqnarray}
\Sigma_{nm}&=&\Bigl\langle\Bigl({\delta H_n-\delta H_{n-1}\over H_0}\Bigr)
                          \Bigl({\delta H_m-\delta H_{m-1}\over H_0}\Bigr)
\Bigr\rangle\nonumber\\
&=&\Bigl\langle\Bigl({\delta H_n^{(v)}-\delta H_{n-1}^{(v)}\over H_0}\Bigr)
               \Bigl({\delta H_m^{(v)}-\delta H_{m-1}^{(v)}\over H_0}\Bigr)
\nonumber\\
&+&\Bigl\langle
   \Bigl({\delta H_n^{(\epsilon)}-\delta H_{n-1}^{(\epsilon)}\over H_0}\Bigr)
   \Bigl({\delta H_m^{(\epsilon)}-\delta H_{m-1}^{(\epsilon)}\over H_0}\Bigr)
\Bigr\rangle\nonumber\\
&=&\Omega_0^{1.2}\int d^3k\,
\Bigl[W^i_n({\bf k})-W^i_{n-1}({\bf k})\Bigr]\hat k_i\cdot\nonumber\\
&&\Bigl[{W^j_m}^*({\bf k})-{W^j_{m-1}}^*({\bf k})\Bigr]\hat k_j
\, {P(k)\over k^2}\nonumber\\
&+&\delta_{nm}
\Bigl\langle\Bigl({\delta H_{n-1}^{(\epsilon)}\over H_0}\Bigr)^2
-\Bigl({\delta H_n^{(\epsilon)}\over H_0}\Bigr)^2\Bigr\rangle,
\label{diff}
\end{eqnarray}
where $\delta_{nm}$ is the Kroeneker $\delta$ function.
Thus, given $\Delta H_n=\delta H_n-\delta H_{n-1}$ from a real sample $D$,
the probability of a critical-density CDM model with a particular set
of $\Gamma$ and $n_{\rm ps}$ is
\begin{equation}
P\langle M\vert DI\rangle\propto {1\over\vert\Sigma\vert^{1/2}}
\exp\Big[-{1\over 2}({\Delta H})^T(\Sigma)^{-1}(\Delta H)\Bigr].
\label{likelihood}
\end{equation}

\begin{figure}
\psfig{figure=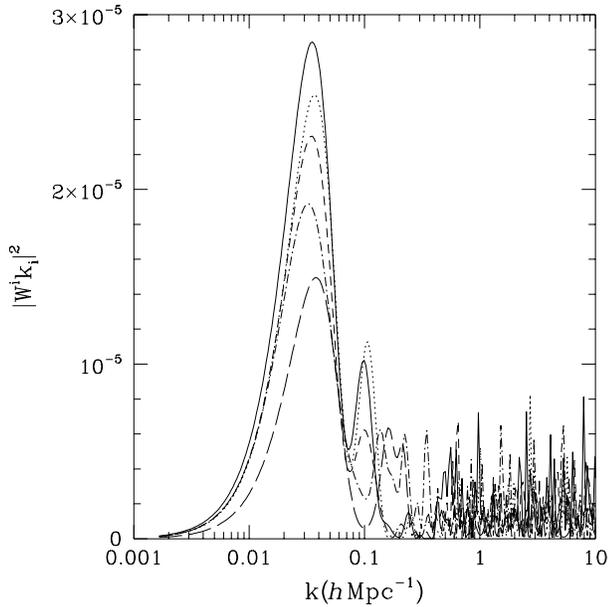,height=3.6in,angle=0}
 \caption{$\vert W^ik_i\vert^2$ of the cluster sample in question.
  Different curves are for different directions in the {\bf k}-space.}
\end{figure}

\section[]{Results}

The formalism is applied to the cluster sample that yields the
template Tully-Fisher relation in the Mark III catalogue
(Willick et al. 1997).  Only subsamples with no less than
15 clusters are included in calculations, because smaller subsamples
are susceptible to large effects of non-linearities.  This cut-off
does not significantly affect the final results due to the large
uncertainties of $\delta H$ in the small subsamples.  The integrals
over the window functions are done by sampling about 40000 ${\bf k}$'s
in space.  The window functions are truncated at $\pi/k\approx 6h^{-1}$ Mpc.
To include the effect of non-linearities below the truncation scale,
a $\sigma_*=300$ km/sec is added
quadratically to the uncertainty in distance of each cluster to fully
represent the uncertainty of its radial peculiar velocity.
The truncation scale is verified not to significantly influence our final
result, but the magnitude of $\sigma_*$ does have an appreciable impact.
A smaller $\sigma_*$ will increase the contrast of the probability distribution
$P\langle M\vert DI\rangle$, thus tighten the resultant constraints.  At the
moment, a $\sigma_*$ of 300 km/sec is quite a safe assumption to make
(Bahcall and Oh 1996).

The test is also applied to the sample of 20 Type Ia SNe of Riess, Press
and Kirshner (1996).  Subsamples run from the 6 nearest Type Ia SNe to the
full sample of 20 SNe.  Truncation of window functions is made at
$\pi/k\approx 4h^{-1}$ Mpc.  The host galaxies of the SNe are assumed to
have a $\sigma_*=480$ km/sec.  This corresponds to a r.m.s. gaussian random
motion of 830 km/sec for host galaxies, which
roughly matches the velocity dispersion of galaxies in clusters.  It is
also the maximal $\sigma_*$ allowed by the sample, because any addition
of distance uncertainties will result in a dispersion of
radial peculiar velocities that is larger than that of data.
Therefore, the constraint we obtain will be rather conservative.

\begin{figure}
\psfig{figure=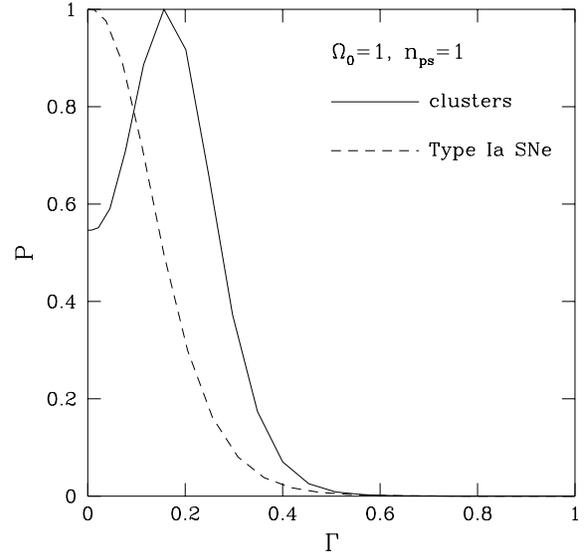,height=3.8in,angle=0}
\caption{Probability distribution of $\Gamma$ when $\Omega_0=1$ and
$n_{\rm ps}=1$, for the cluster sample (the solid curve),
and the Type Ia SN sample (the dashed curve).}
\end{figure}

Figure 2 shows the probability distribution (with an arbitrary normalization)
of $\Gamma$ when $\Omega_0=1$ and $n_{\rm ps}=1$, for the two samples.
After the same distribution function is calculated for other
$\Omega_0$ (in which case $\Lambda=1-\Omega_0$) and $n_{\rm ps}$, 
we find that at 95$\%$ C.L., the cluster sample yields
$\Gamma\le 0.30-0.88(n_{\rm ps}-1)+1.9(n_{\rm ps}-1)^2$, with best fits
$\Gamma = 0.15-0.39(n_{\rm ps}-1)+0.37(n_{\rm ps}-1)^2$; the Type Ia SN sample
yields $\Gamma\le 0.25-0.80(n_{\rm ps}-1)+1.6(n_{\rm ps}-1)^2$ at $95\%$ C.L.,
strongly favoring lower $\Gamma$'s.  These results are accurate for
critical-density CDM models in the range $0.3\le\Omega_0\le 1$ to
$\approx 10\%$, because the factor $\Omega_0^{0.6}$ in peculiar velocities
is mostly offset by the $\Omega_0$ dependence in the COBE normalization
eq.~(\ref{normalize}).  The results indicate that a critical-density
matter-dominated universe with $H_0\ga 50$ km/sec/Mpc
and $n_{\rm ps}\ga 0.8$ cannot fit the variations of Hubble flows in the data.
(The constraints will not change significantly if one includes the tensor
contribution to the COBE normalization of $P(k)$, but will tend to relax
when one decreases $P(k)$ on $\sim 100h^{-1}$ Mpc scales.)
With the number of Type Ia SNe with distance measurements
doubling very soon, we can only expect that the limit
on $\Gamma$ becomes more secure and interesting.

\section*{Acknowledgments}

The author thanks U. Pen for helpful suggestions.
The work is supported by grants NASA NAG5-3062 and NSF PHY95-03384 at UCSD.

\end{document}